\documentstyle[aps,prl,epsf,twocolumn,floats]{revtex}

\renewcommand{\epsilon}{\varepsilon}

\begin{document}
\draft

\twocolumn[\hsize\textwidth\columnwidth\hsize\csname@twocolumnfalse\endcsname

\title{Nonmonotonical crossover of the effective susceptibility exponent}
\author{Erik Luijten\cite{email} and Henk~W.~J. Bl\"ote}
\address{Department of Physics, Delft University of Technology,
         Lorentzweg 1, 2628 CJ Delft, The Netherlands}
\author{Kurt Binder}
\address{Institut f\"ur Physik, WA 331, Johannes Gutenberg-Universit\"at,
  D-55099 Mainz, Germany}
\date{\today}
\maketitle

\begin{abstract}
  We have numerically determined the behavior of the magnetic susceptibility
  upon approach of the critical point in two-dimensional spin systems with an
  interaction range that was varied over nearly two orders of magnitude.  The
  full crossover from classical to Ising-like critical behavior, spanning
  several decades in the reduced temperature, could be observed. Our results
  convincingly show that the effective susceptibility exponent~$\gamma_{\rm
    eff}$ changes {\em nonmonotonically\/} from its classical to its Ising
  value when approaching the critical point in the ordered phase. In the
  disordered phase the behavior is monotonic.  Furthermore the hypothesis that
  the crossover function is universal is supported.
\end{abstract}

]
\narrowtext

At a continuous phase transition several thermodynamic observables diverge as a
power of the temperature distance to the critical point.  These powers, or
critical exponents, have universal values which are identical for large classes
of systems. For example, uniaxial ferromagnets, binary alloys, simple fluids,
binary mixtures, ionic solutions, and polymer mixtures all belong to the
three-dimensional Ising universality class. However, the corresponding
power-law behavior is only observed asymptotically close to the critical point.
As stated by the Ginzburg criterion~\cite{ginzburg}, classical or
mean-field-like critical behavior may be observed at temperatures further away
from the critical temperature~$T_{\rm c}$. The explanation of this {\em
  crossover\/} in terms of competing fixed points of a renormalization-group
transformation is one of the great achievements of Wilson's renormalization
theory.  Nevertheless, the precise nature of the crossover between these two
universality classes is still subject to debate.  Theoretically several
attempts have been made to approximately calculate crossover functions.  For
instance, Nicoll and Bhattacharjee~\cite{nicoll81} solved the renormalization
equations in $d$~dimensions to second order in~$\epsilon=4-d$ by applying a
specific matching condition, whereas Bagnuls and Bervillier~\cite{bagnuls85}
used massive field theory in $d=3$. The results of Belyakov and
Kiselev~\cite{belyakov} are phenomenological generalizations of first-order
$\epsilon$-expansions. All these results are only valid in the symmetric phase
($T>T_{\rm c}$) and suggest that the crossover behavior is universal. However,
Anisimov {\em et al.\/}~\cite{anisimov92} claimed that, while at criticality
microscopic cutoff effects may be neglected compared to the infinite
correlation length, this is no longer the case in the crossover region.  This
implies that the crossover functions cannot be represented as universal
functions of one variable.  A particular question concerns the variation of the
so-called effective exponents describing the continuous change from one type of
power-law behavior to another in the crossover region.  Whereas some
calculations predict a strictly monotonical variation, others indicate that a
{\em nonmonotonical\/} variation might be possible. While on the theoretical
side several important open questions remain, the experimental situation is
hardly better.  Measurements in the critical region are difficult and accurate
results are scarce.  Fisher~\cite{fisher-eff} has discussed experiments on
micellar solutions (expected to belong to the Ising universality
class)~\cite{corti} that yielded values for the susceptibility
exponent~$\gamma$ that lie below the classical value $\gamma_{\rm MF}=1$, while
the Ising value is given by $\gamma_{\rm I}=1.237$~\cite{ic3d}.  He argues that
these results can be incorporated in a standard scaling description of
crossover behavior if one allows for an effective susceptibility exponent that
varies nonmonotonically as a function of the reduced temperature $t = (T-T_{\rm
  c})/T_{\rm c}$.  More recently, Anisimov {\em et al.\/}~\cite{anisimov95}
measured a susceptibility~$\chi$ for which they found that the logarithmic
derivative $\gamma_{\rm eff} \equiv - d\ln \chi / d\ln |t|$ approached the
Ising value from {\em above\/} upon approach of the critical point. As this
implies a nonmonotonical variation of~$\gamma_{\rm eff}$, Bagnuls and
Bervillier subsequently suggested that the measurements might have been taken
outside the critical region, see Refs.~\cite{bagnuls96,anisimov96}.  Indeed,
since the crossover region is expected to span several decades in the reduced
temperature~\cite{anisimov92,fisher-eff}, in many experiments the full
crossover behavior cannot be observed. At the same time, the large extent of
the crossover region reinforces its experimental relevance: many measurements
of critical exponents are actually made {\em within\/} the crossover region and
thus only a detailed knowledge of the crossover behavior guarantees a correct
interpretation of the data.

Although it is tempting to apply numerical simulations to shed some light on
these issues, in practice one encounters difficulties comparable to those
experienced by experimentalists. In particular the size of the crossover region
constitutes a towering hurdle.  A major effort has been undertaken in
Ref.~\cite{deutsch93} for three-dimensional polymer mixtures, in which
crossover occurs as a function of the polymer chain length. These systems offer
the advantage that the crossover can be influenced both by varying the
temperature and by changing the chain length. Despite chain lengths of up to
512~monomers, the results did not span the full crossover region. Mon and
Binder~\cite{mon} examined the two-dimensional Ising model with an extended
range of interaction~$R$, where crossover from Ising to classical behavior
occurs when $R$ is increased. They studied crossover in finite systems at
$T=T_{\rm c}$.  Even in these systems the mean-field regime could hardly be
reached. In Ref.~\cite{medran} we showed that a new Monte Carlo~(MC) cluster
algorithm for long-range interactions~\cite{ijmpc} could be applied to this
model, leading to a speed increase of many orders of magnitude compared to
conventional algorithms.

In this Letter we use this algorithm to study two-dimensional Ising systems
with a variable interaction range and present results for the crossover
behavior of the magnetic susceptibility at temperatures below and above~$T_{\rm
  c}$.  Although two-dimensional systems are simpler than their
three-dimensional counterparts, this model exhibits a surprising behavior. In
particular, a qualitative difference between $T<T_{\rm c}$ and $T>T_{\rm c}$ is
found.  The advantage of examining two-dimensional instead of three-dimensional
systems is the much larger variation of the critical exponents in the crossover
region and the accessibility of larger interaction ranges, which makes it
feasible to cover the full crossover region.

The model under investigation was introduced in Ref.~\cite{mon} and is defined
by the following Hamiltonian,
\begin{equation}
{\cal H}/k_{\rm B}T = - \sum_{ij} K_d({\bf r}_i - {\bf r}_j)s_i s_j \;,
\end{equation}
where the spins~$s$ take the values~$\pm 1$, the sum runs over all spin pairs,
and the spin--spin coupling depends on the distance $|{\bf r}|$ between the
spins as $K_d({\bf r})=cR_m^{-d}$ for $|{\bf r}| \leq R_m$ and $K_d({\bf r})=0$
for $|{\bf r}|>R_m$. For finite~$R_m$ the critical behavior of this model will
be Ising-like, but for $R_m \to \infty$ it will be classical. This implies a
singular dependence of the critical amplitudes on~$R_m$, which was first
derived on phenomenological grounds in Ref.~\cite{mon}. In Ref.~\cite{medran} a
renormalization derivation of these singular dependences was given, which in
addition revealed logarithmic corrections for $d=2$. To avoid lattice effects
we formulate range dependences in terms of an effective interaction range~$R$,
which is directly related to~$R_m$~\cite{mon}.  The Ginzburg criterion
introduces a parameter $G \propto R^{-2d/(4-d)}$ (the Ginzburg number) which
determines whether the critical behavior will be Ising-like ($t \ll G$) or
classical ($t \gg G$). In the latter case, care must be taken that $t$ is still
within the critical region. For many experimental systems the Ginzburg number
is not small and one has left the critical region before observing the full
crossover to classical critical behavior. In our model system, $G$ is
adjustable so that we can vary $t/G$ over the full crossover region while
keeping $t$ sufficiently small.  On the other hand, for a too small Ginzburg
number the critical point must be approached very closely to access the Ising
regime. The diverging correlation length is then in our simulations truncated
by the finite system size~$L$. Therefore we construct the crossover function by
studying systems with various values of~$G$ (interaction ranges) such that $t$
has to be varied only within a limited range (but in such a way that the
results for several different~$G$ overlap at fixed~$t/G$).

We have carried out MC simulations of square systems with periodic boundary
conditions containing up to $1000 \times 1000$ spins in which each spin
interacts with up to $31416$ neighbors.  This corresponds to an effective
interaction range~$R$ of $71$~lattice spacings or intermolecular distances.  To
avoid systematic errors in the determination of the crossover behavior, an
accurate estimate of~$T_{\rm c}$ as a function of~$R$ is required. For systems
with interaction ranges up to $R \approx 8.3$ results for $T_{\rm c}(R)$ have
been obtained in Ref.~\cite{medran}. For larger ranges the critical temperature
can be calculated to a comparable accuracy from a renormalization expression
for $T_{\rm c}(R)$~\cite{medran}. Further simulational details will be
presented elsewhere~\cite{cross}.

In the two-dimensional Ising model $\gamma_{\rm I}=7/4$ and the susceptibility
$\chi$ diverges for $t \uparrow 0$ as $A_{\rm I}^{-}(-t)^{-7/4}$ and for $t
\downarrow 0$ as $A_{\rm I}^{+}t^{-7/4}$. The critical amplitudes for the
nearest-neighbor model are known exactly~\cite{barouch}, $A_{\rm
  I}^{-}=0.025537\ldots$ and \@$A_{\rm I}^{+}=0.96258\ldots\,$. Note the very
large asymmetry, $A_{\rm I}^{+}/A_{\rm I}^{-} \approx 38$.  Mean-field theory
predicts a susceptibility that for $t \uparrow 0$ diverges as $1/(-2t)$ and for
$t \downarrow 0$ as $1/t$, i.e.\ a susceptibility exponent $\gamma_{\rm MF}=1$
and a much smaller ratio $A_{\rm MF}^{+}/A_{\rm MF}^{-}=2$.  As derived in
Refs.~\cite{mon,medran}, the Ising critical amplitude of the susceptibility is
proportional to $R^{-3/2}$. Thus, in a graph displaying the results for various
ranges as a function of the crossover variable $t/G \propto tR^2$ a data
collapse is obtained for $\chi/R^2$. The susceptibility is related to the
average magnetization per spin~$m$. In our simulations we have for $t<0$
sampled the connected susceptibility given by the fluctuation relation
$\tilde{\chi} = L^d(\langle m^2 \rangle - \langle |m| \rangle^2)/ k_{\rm B}T$,
whereas for $t>0$ we have used $\chi = L^d \langle m^2 \rangle /k_{\rm B}T$.

\begin{figure}[tbp]
\begin{center}
\leavevmode
\epsfxsize 3.09in
\epsfbox{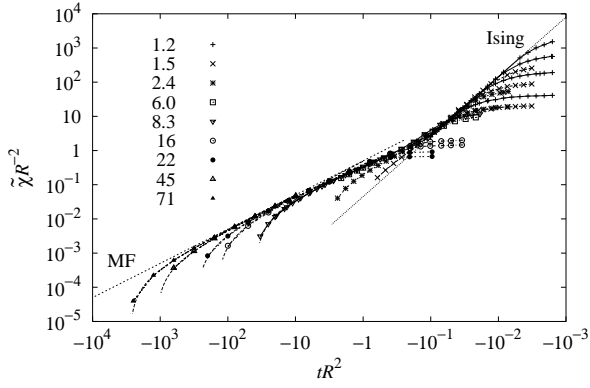}
\end{center}
\caption{Crossover behavior of the connected susceptibility~$\tilde{\chi}$ for
  various ranges and system sizes. In this and all following figures the
  numbers in the key refer to values for the interaction range~$R$.}
\label{fig:chi-low}
\end{figure}
\begin{figure}[hbtp]
\begin{center}
\leavevmode
\epsfxsize 3.09in 
\epsfbox{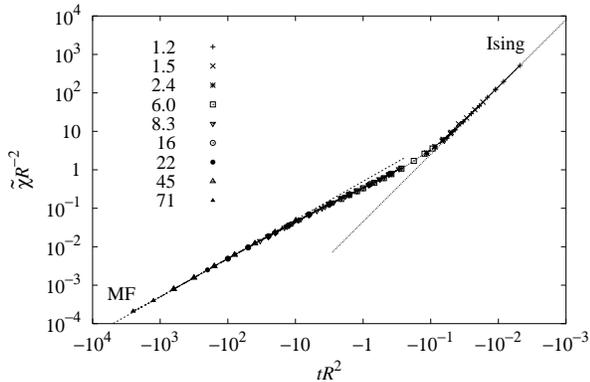}
\end{center}
\caption{Crossover curve for the connected susceptibility~$\tilde{\chi}$.}
\label{fig:chi-corr}
\end{figure}
\begin{figure}[hbpt]
\begin{center}
\leavevmode
\epsfxsize 3.09in
\epsfbox{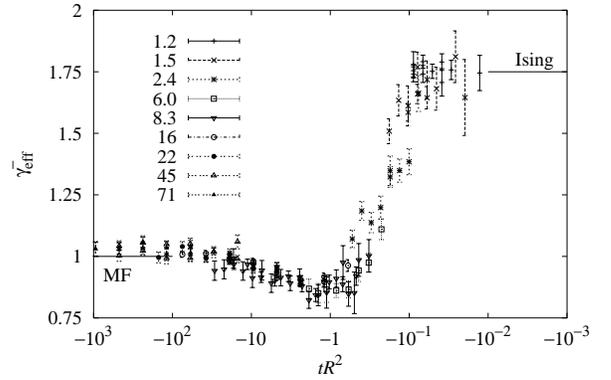}
\end{center}
\caption{The effective susceptibility exponent~$\gamma^{-}_{\rm eff}$ below
  $T_{\rm c}$.}
\label{fig:gammalow}
\end{figure}

In Fig.~\ref{fig:chi-low} we show the magnetic susceptibility below~$T_{\rm c}$
for various system sizes and interaction ranges. This graph exhibits several
interesting features. For very small values of $|t|$ the curves lie almost
horizontal; this is the finite-size regime where the correlation length is
truncated by the system size. For somewhat larger values of~$|t|$ the curves
start following the Ising asymptote with slope (i.e., the logarithmic
derivative $d\ln \chi / d\ln |t|$)~$-7/4$. This is the critical behavior as it
is experimentally measured close to~$T_{\rm c}$. Also the critical
amplitude~$A_{\rm I}^{-}$ is accurately reproduced by the simulations.  At even
lower~$T$, we see that the curves for systems with small interaction ranges
start to deviate from the Ising asymptote toward the mean-field asymptote
(slope~$-1$) without actually reaching it.  These systems have left the
critical region and the order parameter shows strong saturation effects (which
decreases the susceptibility). However, systems with a larger interaction range
clearly cross over to the mean-field asymptote, reproducing the mean-field
critical amplitude~$A_{\rm MF}^{-}$. For even lower temperatures these systems
also exhibit saturation effects, which for~$R \gtrsim 8.3$ are accurately
described by mean-field theory (dashed curves).  However, the outstanding
feature of this graph is the region between the Ising and the mean-field
asymptote. Namely, before settling at the latter asymptote, the curve
describing the susceptibility first has (in this double-logarithmic plot) a
slope that is {\em less steep\/} than in the mean-field regime. That is,
$\gamma^{-}_{\rm eff} < 1$ (the superscript minus sign indicates that we are
considering the case $t<0$). To illustrate this effect more clearly we have
reproduced Fig.~\ref{fig:chi-low} without the data that are plagued by
finite-size effects or lie outside the critical region.  Furthermore, we have
corrected for the saturation effects for systems with large ranges. While this
is a real physical effect, it can be removed by applying a correction factor
accounting for the difference between the asymptotic mean-field susceptibility
and the mean-field susceptibility affected by saturation.  The resulting graph
is shown in Fig.~\ref{fig:chi-corr}.
The nonmonotonical variation of the slope is now clearly visible. The data for
different interaction ranges $1 \lesssim R \lesssim 70$ overlap for
considerable intervals of~$tR^2$. The perfect collapse of these data lends
strong support to the hypothesis that the crossover curve is universal and
spans several decades in the reduced temperature. In addition, it follows from
the renormalization treatment in Ref.~\cite{medran} that the correlation
length~$\xi$ decreases as~$t^{-\nu}$ with an amplitude which is for $d=2$ to
leading order independent of~$R$. Thus, at a fixed value of~$tR^2$ the curves
for different ranges have {\em different\/} values for~$\xi$ and the fact that
they collapse implies that the ratio between $\xi$ and the lattice spacing~$a$
does not affect the crossover curve. This is markedly different from the
results of Ref.~\cite{anisimov92} for $d=3$. Also the influence of irrelevant
fields is not visible in the data collapse.  To connect to experimental
results, we have plotted the effective exponent~$\gamma^{-}_{\rm eff}$
(obtained by numerical differentiation) in Fig.~\ref{fig:gammalow}.  Starting
from~$T_{\rm c}$, $\gamma^{-}_{\rm eff}$ first steeply decreases to a minimum
below~$\gamma_{\rm MF}$ and then gradually rises to the asymptotic mean-field
value.

\begin{figure}[b]
\begin{center}
\leavevmode
\epsfxsize 3.09in
\epsfbox{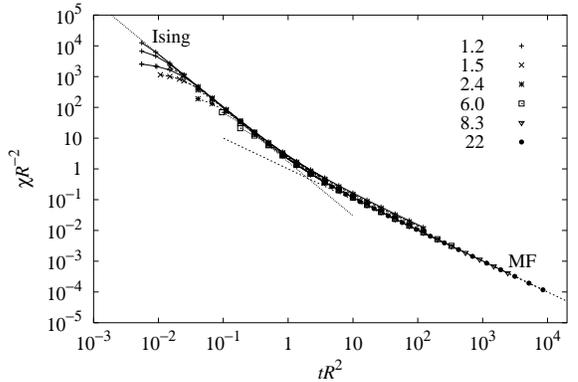}
\end{center}
\caption{Crossover behavior of the susceptibility~$\chi$ for various ranges
  and system sizes.}
\label{fig:chi-high}
\end{figure}

\begin{figure}[htbp]
\begin{center}
\leavevmode
\epsfxsize 3.09in
\epsfbox{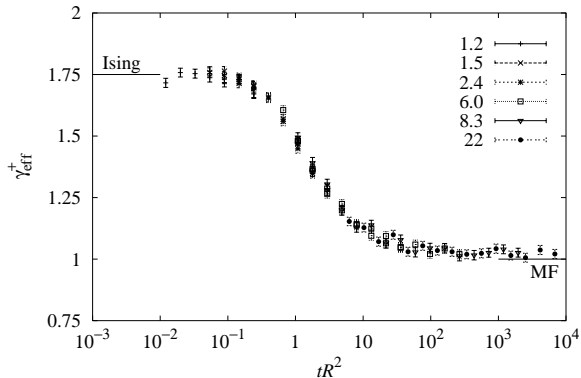}
\end{center}
\caption{The effective susceptibility exponent~$\gamma^{+}_{\rm eff}$ above
  $T_{\rm c}$.}
\label{fig:gammahigh}
\end{figure}

Now we turn to the symmetric phase, for which a data collapse of the
susceptibility~$\chi$ is shown in Fig.~\ref{fig:chi-high}.
Just as below~$T_{\rm c}$, finite-size effects occur for very small~$t$.
Outside the finite-size regime the data for various~$R$ nicely collapse on the
Ising asymptote, again with slope~$-7/4$ but with a much larger amplitude. For
higher temperatures the curves appear to gradually approach the mean-field
asymptote.  However, only for larger interaction ranges the critical
amplitude~$A_{\rm MF}^{+}$ is reproduced. This stresses an important point:
above~$T_{\rm c}$ no saturation of the order parameter occurs, marking the end
of the critical region, but the system smoothly passes over to regular
(noncritical) behavior.  In this high-temperature region the susceptibility
decreases proportional to~$1/T$. It is this behavior that is seen in the graph
at high temperatures for systems with small interaction ranges. As rightfully
stressed by Bagnuls and Bervillier~\cite{bagnuls96}, this behavior should be
clearly distinguished from classical {\em critical\/} behavior. Thus, it is by
no means disturbing that the curves for small~$R$ deviate from the mean-field
asymptote and this does not imply a nonuniversal character of the crossover
curve in the critical region.  Disregarding these systems that have left the
critical region, we note that above~$T_{\rm c}$ the susceptibility smoothly
crosses over from Ising-like to classical critical behavior and that the
effective exponent~$\gamma^{+}_{\rm eff}$ decreases monotonically from~$7/4$
toward~$1$, as visualized in Fig.~\ref{fig:gammahigh}.

In conclusion, we have presented crossover curves for the magnetic
susceptibility in two-dimensional Ising models with medium-range interactions,
both for $t<0$ and~$t>0$.  Unlike previous treatments, which all suffered from
some systematic limitations (like extrapolating low-order $\epsilon$-expansions
to physical dimensions), the present approach for the first time gives an
explicit description of crossover scaling functions for critical phenomena.  At
least in principle, these curves could be directly compared to experimental
results.  The large interaction ranges that could be accessed allowed us to
observe the crossover between Ising-like and classical critical behavior for
nearly six decades in the crossover variable.  This has yielded for the first
time strong numerical evidence that below~$T_{\rm c}$ the effective
susceptibility exponent~$\gamma_{\rm eff}$ varies nonmonotonically between its
Ising value and its classical value.  Above~$T_{\rm c}$, on the other hand, the
exponent shows a monotonical variation between the two limiting values. Thus,
the plausibility of the occurrence of a minimum in~$\gamma_{\rm eff}$ in
three-dimensional systems, at least in the phase of broken symmetry, has been
greatly increased.  Furthermore, the fact that the crossover curves for many
different interaction ranges collapse supports the hypothesis that this curve
is universal.

\acknowledgments
E.~L. and H.~B. acknowledge the kind hospitality of the condensed matter theory
group of the Johannes Gutenberg-Universit\"at Mainz, where part of this work
has been completed. We are grateful to the HLRZ J\"ulich for access to a
Cray-T3E, on which part of the computations have been carried out.







\end{document}